\documentclass[preprint*,  12pt]{JHEP3} 

\JHEPspecialurl{http://jhep.sissa.it/JOURNAL/JHEP3.tar.gz}

\usepackage{epsfig,multicol}

\newcommand\fverb{\setbox\pippobox=\hbox\bgroup\verb}
\newcommand\fverbdo{\egroup\medskip\noindent%
            \fbox{\unhbox\pippobox}\ }
\newcommand\fverbit{\egroup\item[\fbox{\unhbox\pippobox}]}
\newcommand{\lsim}{\mathrel{\mathop{\kern 0pt \rlap
  {\raise.2ex\hbox{$<$}}}
  \lower.9ex\hbox{\kern-.190em $\sim$}}}
\newcommand{\gsim}{\mathrel{\mathop{\kern 0pt \rlap
  {\raise.2ex\hbox{$>$}}}
  \lower.9ex\hbox{\kern-.190em $\sim$}}}
\newcommand{\be}     {\begin{equation}}
\newcommand{\ee}     {\end{equation}}
\newcommand{\bea}     {\begin{eqnarray}}
\newcommand{\eea}     {\end{eqnarray}}
\newbox\pippobox

\title{Implication of Brane fluctuations to
indirect collider signals}

\author{Seong Chan Park \\
    Korea Institute for Advanced Study (KIAS)\\
207-43 Cheongryangri-dong, Dongdaemun-gu Seoul 130-012, Korea\\
Email: \email{spark@kias.re.kr}}
                \author{Hee Sung Song \\
         Center for Theoretical
physics and School of Physics, Seoul National University,\\ Seoul
151-742, Korea \\E-mail: \email{hssong@physs.snu.ac.kr}}

\preprint{\hepph{0109258}\\
 SNUTP-01033 \\
 KIAS-P03025}  

\abstract{We study the effect of brane fluctuation on the indirect
signals of high energy colliders. Brane fluctuation could act as a
regulator of divergent expression of infinite tower of
Kaluza-Klein graviton effects. The phenomenological parameter
$\lambda$, introduced by Hewett, is shown to be determined in our
setting, and its dramatic behaviors depending on the $D=4+\delta$
dimensional gravitation scale $M_D$, `softening parameter'
$\Delta$, and $\sqrt{s}$ of collider are presented. The present
exclusion bounds from the processes $e^+e^-\rightarrow
\gamma\gamma$ and $p\bar{p}\rightarrow e^+ e^-,
\gamma\gamma(\gamma)$ are considered within the parameter space
$(M_D, \Delta)$ with respect to the number of extra dimensions.}

\begin{document}


\section{Introduction}
It is phenomenologically interesting when the size of extra
dimension is so large that the Kaluza-Klein excitations of bulk
fields could  directly affect the low energy phenomena . If all
the standard model(SM) fields are confined on the brane and only
graviton can propagate through the bulk, the size of extra
dimension can be large enough to nullifying the hierarchy between
the weak scale and the Planck scale
\cite{ADD},\cite{Antoniadis:1998ig}. In that case, the
Kaluza-Klein (KK) tower of graviton can make
 sizable contributions to the collider physics \cite{Han:1998sg},\cite{Giudice:1998ck}.
Not only signals with direct production of KK graviton, but
indirect signals with the KK mediation can provide chances to
detect the effects from the extra dimensions. The scattering cross
section including KK graviton mediating diagrams can be written as
\be \sigma_{\rm Total}= \sigma_{\rm SM} + \eta_{KK} \sigma_{\rm
Mix}
                    + \eta_{KK}^2 \sigma_{KK}.
\ee
Where $\eta_{KK}$ denotes the propagating factor of KK graviton
tower defined as
\be \eta_{KK}\equiv \frac{i^2}{8 M_P^2}\sum_{KK}
\frac{1}{s-m_{KK}^2} \ee
when only s-channel diagrams are involved. Here $m_{KK}^2 \equiv
\vec{n}\cdot\vec{n}/R^2 $ is the mass of the KK graviton in state
$\vec{n}=(n_1, n_2 ,...,n_\delta)$, the factor $1/8$ is introduced
for future convenience and the Planck scale $M_P$ is for
gravitational coupling. The case with t-, u- channel KK mediating
diagrams also contributing like $e^+e^-\rightarrow e^+e^-$ can be
considered as a straightforward ways. The summation through the
whole tower of KK states is generally divergent and so we should
take a cutoff scale ($N_\Lambda =\frac{\Lambda}{R}$) to get finite
result and then we simply throw the rests away \cite{Sundrum}
(see, for loop calculations of Kaluza-Klein states
\cite{Strumia},\cite{Giudice:2003tu}). The size of extra dimension
$R$ is related to the scales of the gravitation as $R^\delta
M_D^{2+\delta} = M_P^2$, where $\delta$ is number of extra
dimensions. With small spacing ($\sim \frac{1}{R}$) the summation
could be approximated to the integration, we can obtain
$\eta_{KK}$ at the limit $\Lambda/\sqrt{s} >>1$ as
\bea \eta_{KK}(Rigid)~~
          \approx  \frac{\pi}{2 M_D^4}\ln(\Lambda/\sqrt{s})
     ~~~~~~~&&(\delta=2), \\
      \approx  \frac{\Omega_\delta s^{\delta/2-1}}{8 M_D^{\delta+2}}
             (\Lambda/\sqrt{s})^{\delta-2} ~~&&(\delta>2)
\eea where $\Omega_\delta$ is the solid angle in $\delta$
dimensional space, e.g., $\Omega_2 = 2\pi$ in 2 dimensional space.

By taking $\Lambda \sim M_D$, the factor could be estimated as
$\eta_{KK}\sim \lambda/M_D^4$, where $\lambda$ encapsulates all
the uncertainties from the number of extra dimensions, the unknown
relation between $\Lambda$ and $M_D$, and threshold effects coming
from the string theory beyond the cutoff scale \cite{Hewett}.
Though it is obvious that $\lambda$ is dependent on the number of
extra dimensions and the energy scale of collider $\sqrt{s}$, it
is usually assumed that the value is ${\cal O}(1)$ and insensitive
to the number of extra dimensions. With this assumptions, indirect
signals of extra dimensions have been treated independently of
$\delta$ etc. \cite{Review}, \cite{Review2}. However from a more
close study including the dynamics of the brane fluctuation,
$\lambda$ shows dramatic behavior depending $\delta$, $\sqrt{s}$
and softening scale parameter which is essentially determined by
tension of the brane \cite{Yan:2001tg}, \cite{Alcaraz:2002iu}.
This paper is organized as following. In the sec.II, we first
present the explicit formula for $\eta_{\rm KK}$ in terms of the
brane tension, number of extra dimensions and the C.M. energy of
the colliding particles. Using the formula we can find the
appropriate expression for the phenomenological parameter
$\lambda$ in terms of the tension parameter. From the expressions
we can `translate' the experimental bounds from LEP-II and
Tevatron in terms of brane tension scale and $M_D$ scale in
sec.III. Summary and conclusion of our study will be given in
sec.IV.

\section{Brane Fluctuations and determining $\lambda$ parameter}
In the string theory embedding of large extra dimensional theory,
our world might be a dynamical object carrying finite tension
\cite{String}. It is very natural since any relativistic
consideration does not allow strictly rigid objects. Thus the
formalism including brane fluctuation must ultimately be employed
to probe the high energy physics of extra dimensions with
brane\cite{Sundrum},\cite{Bando}. The brane fluctuation could be
described by introducing Nambu-Goldstone boson $\vec{\phi}(x)$
which came from the spontaneous translational symmetry breaking.
The dynamics of the Nambu-Goldstone
 boson, inducing the metric on the fluctuating brane,
is described by the Nambu-Goto action with the induced metric.
\be {\cal L}=-\tau \int d^4 x \sqrt{-g}= -\tau \int d^4 x
(1-\frac{1}{2}\partial_\mu \vec{\phi}(x) \cdot \partial^\mu
\vec{\phi}(x)
 + \cdots)
\ee
where the tension of the brane is denoted as $\tau$. After
expanding the bulk graviton field around the compact extra
dimensions and taking normal ordering for the expansion modes in
perturbation framework, the interaction Lagrangian with the KK
gravitons and the SM particles is shown to carry an exponential
`softening factor' \cite{Murayama},\cite{Park}:
\be {\cal L} \supset -\frac{1}{M_P}g_{\mu\nu}T^{\mu\nu}({\rm SM})
\Rightarrow
-\frac{1}{M_P}\sum_{\vec{n}}e^{-\frac{1}{2}m_{KK}^2/\Delta^2 }
g_{\mu\nu}^{\vec{n}}T^{\mu\nu}({\rm SM}) \ee
where $\frac{1}{\Delta^2}$ is the free propagator of $\vec{\phi}$,
\be \frac{1}{\Delta^2} \equiv <\phi(x)\phi(y)>|_{x\rightarrow y}
=-\frac{1}{4\pi^2 \tau} (x-y)^{-2}|_{x\rightarrow y} \cite{Bando}.
\ee
In principle, the scale $\Delta^2$ could be determined by the
tension of the brane and the cutoff scale of loops of a $\phi(x)$
field. In this study we just take the scale $\Delta$ as a free
parameter of the theory which shows the effect of the brane
fluctuation. Note that when the scale is chosen at infinity, the
action reduced to the normal UN-fluctuating case and if it is
chosen at the same order of $M_D$ scale it will provide very rich
phenomenology of high energy colliders. Since the coupling of the
higher KK states are quite suppressed with the exponential
softening factor, divergent expression for the infinite KK
graviton contribution could be naturally regularized.

For the indirect collider signals, we can get the regularized
expression for $\eta_{KK}$ or $\lambda$ parameter as
\be \eta_{KK}(Fluctuating) \Rightarrow \frac{-1}{8 M_P^2} \int
d^\delta n \frac{e^{-\vec{n}\cdot\vec{n}/(R^2
\Delta^2)}}{s-\vec{n}\cdot\vec{n}/R^2}. \ee
From the above relation we can derive the expression for the
parameter $\lambda$ as follows.
\be \lambda(Fluctuating)=\frac{-\Omega_\delta}{8} (\frac{ \sqrt{s}
}{M_D})^{\delta-2} {\cal I}(\delta, s/\Delta^2). \ee
The integral function is introduced as
\be {\cal I}(\delta, s/\Delta^2)=\int dx \frac{x^{\delta-1}e^{-x^2
(s/\Delta^2)}}{1-x^2} \ee
and we take the principal value for the singular integral without
pole contribution of KK state. The value for the integration is
very stable with respect to
 the UV cutoff
and we understand the behavior with exponential suppression factor
for the large $x$ region. This is very general behavior from the
brane fluctuation working as physical regularization factor. In
Fig.1,2 and 3, we present the numerical results for the parametric
dependence of $\lambda$ parameter. Note that the sign of $\lambda$
is essentially determined by the ratio of $(s/\Delta^2)$ for given
$\delta$. If the ratio is large enough that the contribution of
the light KK gravitons are quite suppressed, the integral function
surely gives negative sign. In that case, $\lambda$ could have
positive sign. But for the case with the large softening
parameter($\Delta >{\cal O}(1)$TeV), $\lambda$ usually carries
negative sign and we would like to concentrate on this possibility
in great detail.

\smallskip
\FIGURE{\epsfig{file=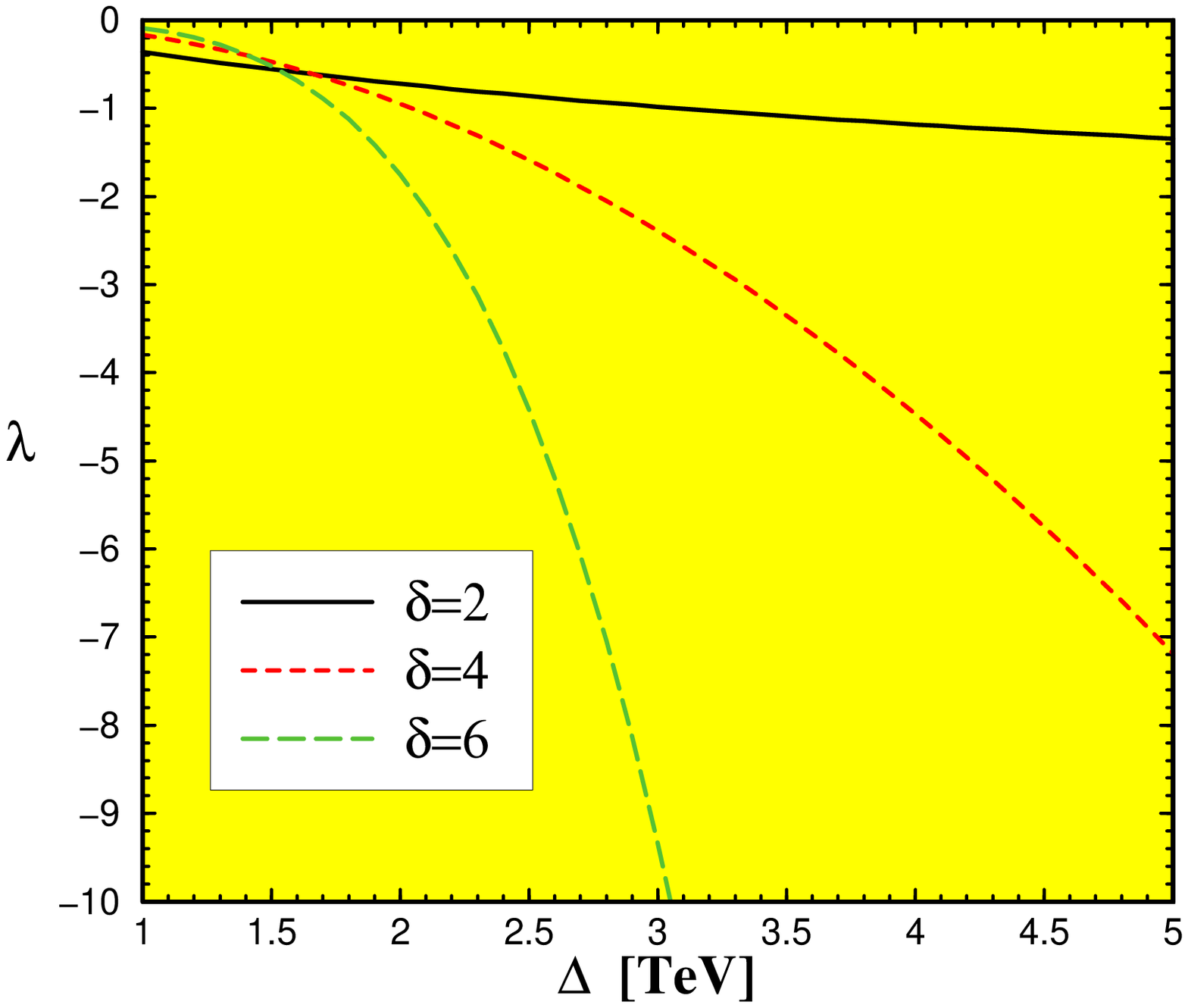,width=6cm,
                    height=6cm}
       \caption{\it
Softening parameter $\Delta$ dependence of $\lambda$. The solid
line, dotted line and dashed line denote the case for $\delta=2,4$
and $6$, respectively.(All the following graphes take the same
convention for $\delta$ lines. $M_D$ and $\sqrt{s}$ are set as 2
and 0.5 TeV, respectively.)}
 \label{fig:fig1}}
(Fig.1) shows the softening parameter dependence of the $\lambda$
parameter when $M_D$ and $\sqrt{s}$ are set to be 2 and 0.5 TeV,
respectively. The $\delta$ dependence is very crucial. It should
be noted that within the reasonable range of the parameter space,
$\lambda$ value is well approximated as ${\cal O}(1)$ value for
the case with $\delta=2$ and $4$. But for the case with
$\delta=6$, the absolute value can be much larger.

In (Fig.2), we plot the $M_D$ dependence when $\Delta$ and
$\sqrt{s}$ are set to be 3 and 0.5 TeV, respectively. The absolute
value is suppressed with larger $M_D$. We can understand this
behavior from the inversely depending relation given above. The
case for $\delta=2$ is very stable with respect to varying $M_D$
but it is not general behavior for larger dimensions. We can also
see that the absolute value is ${\cal O}(1)$ for the case with two
or four extra dimensions but  can be very large when $\delta=6$.
Crossing occurs at $M_D\approx 3$ TeV and which is generally
possible from the non-linear dependence of the parameters.
\smallskip \DOUBLEFIGURE{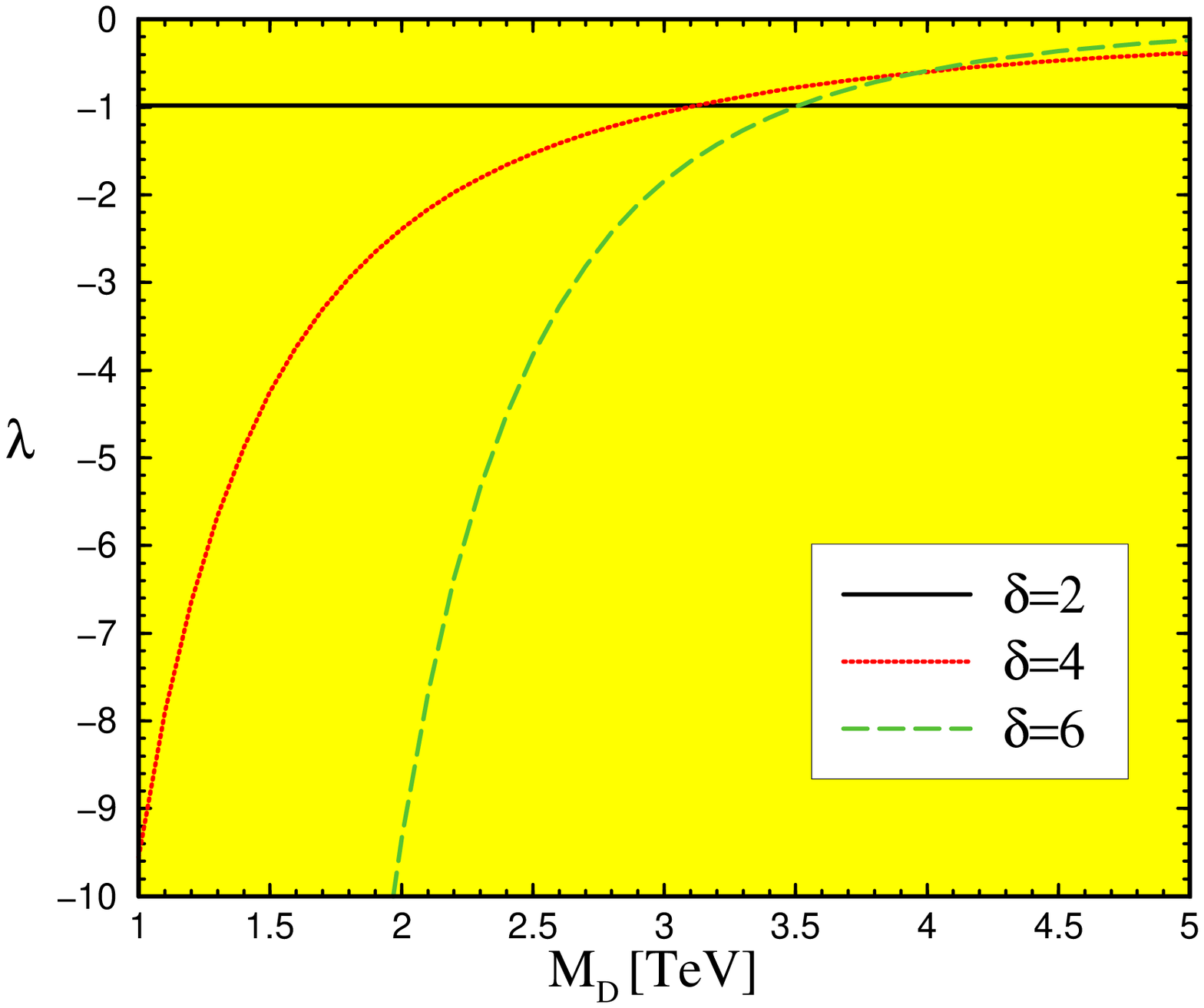,width=6cm,
                    height=6cm}{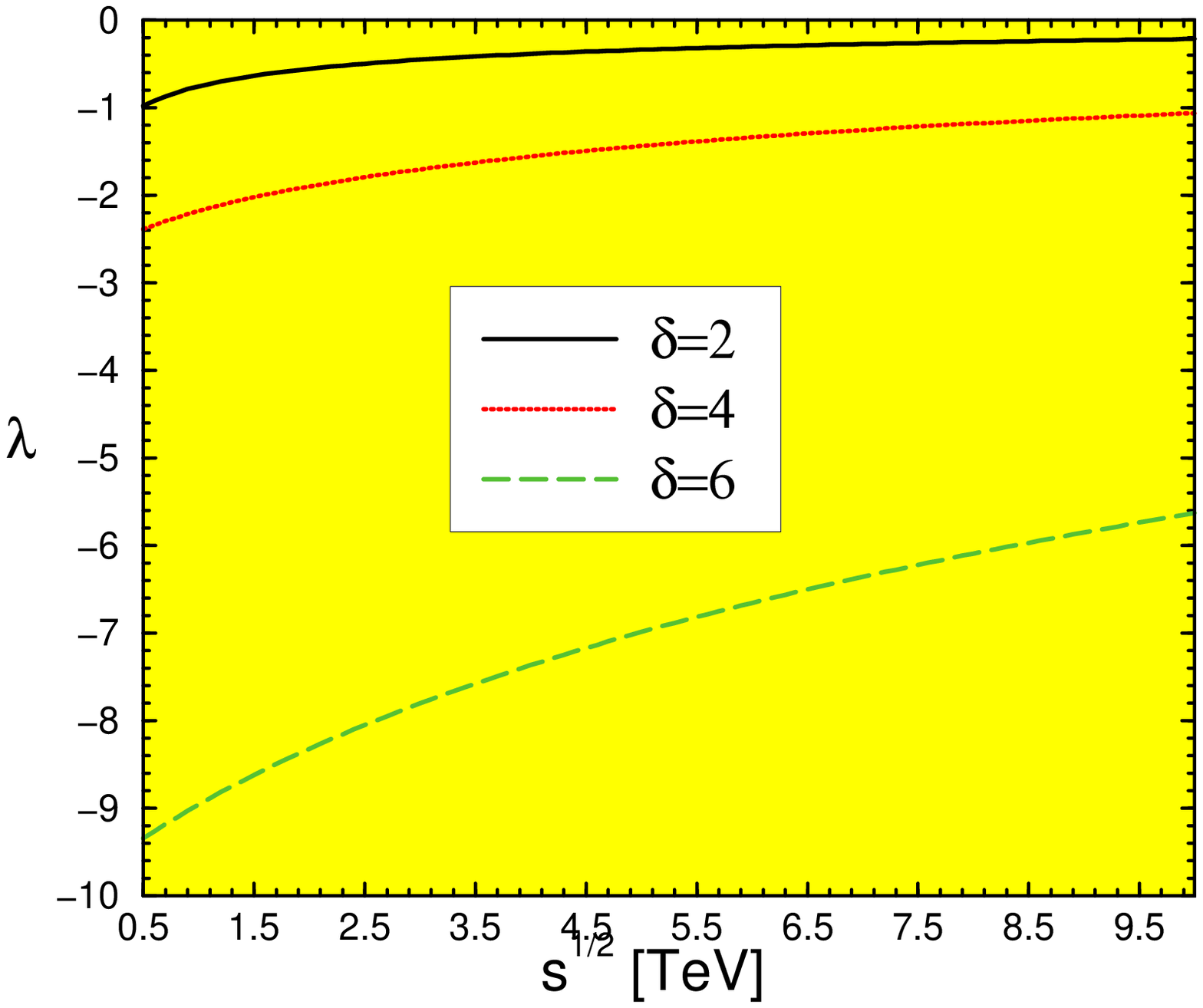,width=6cm,
                    height=6cm}{\it
$M_D$ dependence of $\lambda$ with varying $\delta=2,4$ and 6.
$\Delta$ and $\sqrt{s}$ are set as 3 and 0.5 TeV,
respectively.}{\it The collider C.M. energy $\sqrt{s}$ dependence
of $\lambda$ with respect to the varying $\delta=2,4$ and 6. $M_D$
and $\Delta$ are set as 2 and 3 TeV,respectively.}

In (Fig.3), $\sqrt{s}$ dependence is plotted when $M_D$ and
$\Delta$ are set to be 2 and 3 TeV, respectively. We can see the
rather smooth dependence for two or four extra dimensional cases
but the slope is a bit steeper with six extra dimensions. With
given parameter set, the $\lambda$ has ${\cal O}(1)$ value in the
parameter space.

\section{Collider bounds}
Now let us consider the experimental bounds for the indirect
signals of KK gravitons from the fluctuating brane. Here we
consider the dielectron and diphoton production processes as a
concrete example. However, our method is quite general so that we
can extend the study to any indirect collider signals of extra
dimensions with brane fluctuation regularization. Usually
experimental bounds for the extra dimensions are given by $M_D$
assuming $\lambda= \pm 1$. However,as was seen at the above
section,
 the parameter $\lambda$ generally depends on
not only $M_D$ but also  $\delta$ and $\Delta$ with brane
fluctuation effects. We could only impose experimental bounds
within the parameter space of three dimensions:($M_D, \Delta,
\delta$) with given center of mass energy of collider.

\smallskip \DOUBLEFIGURE{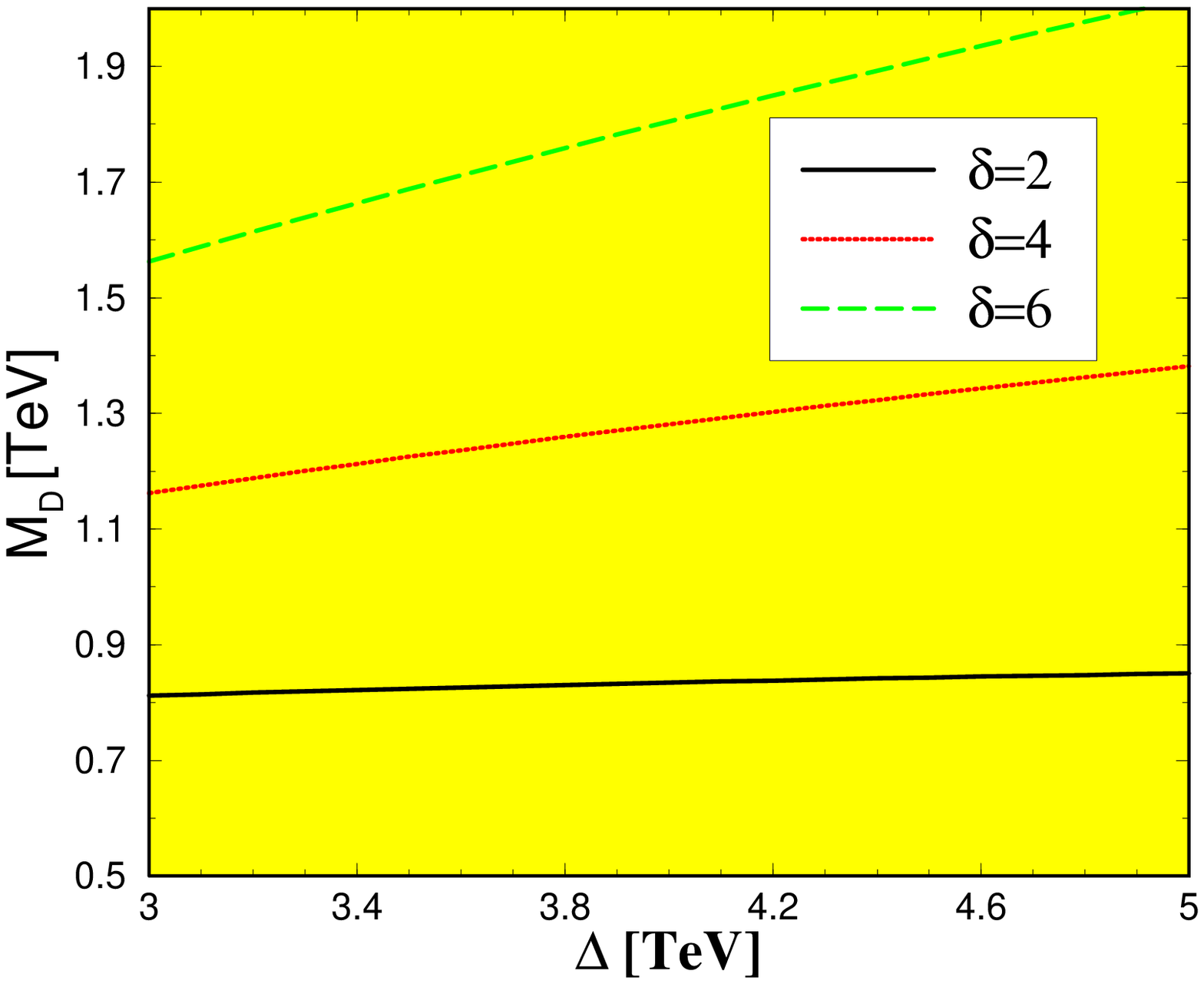,width=6cm,
                    height=6cm}{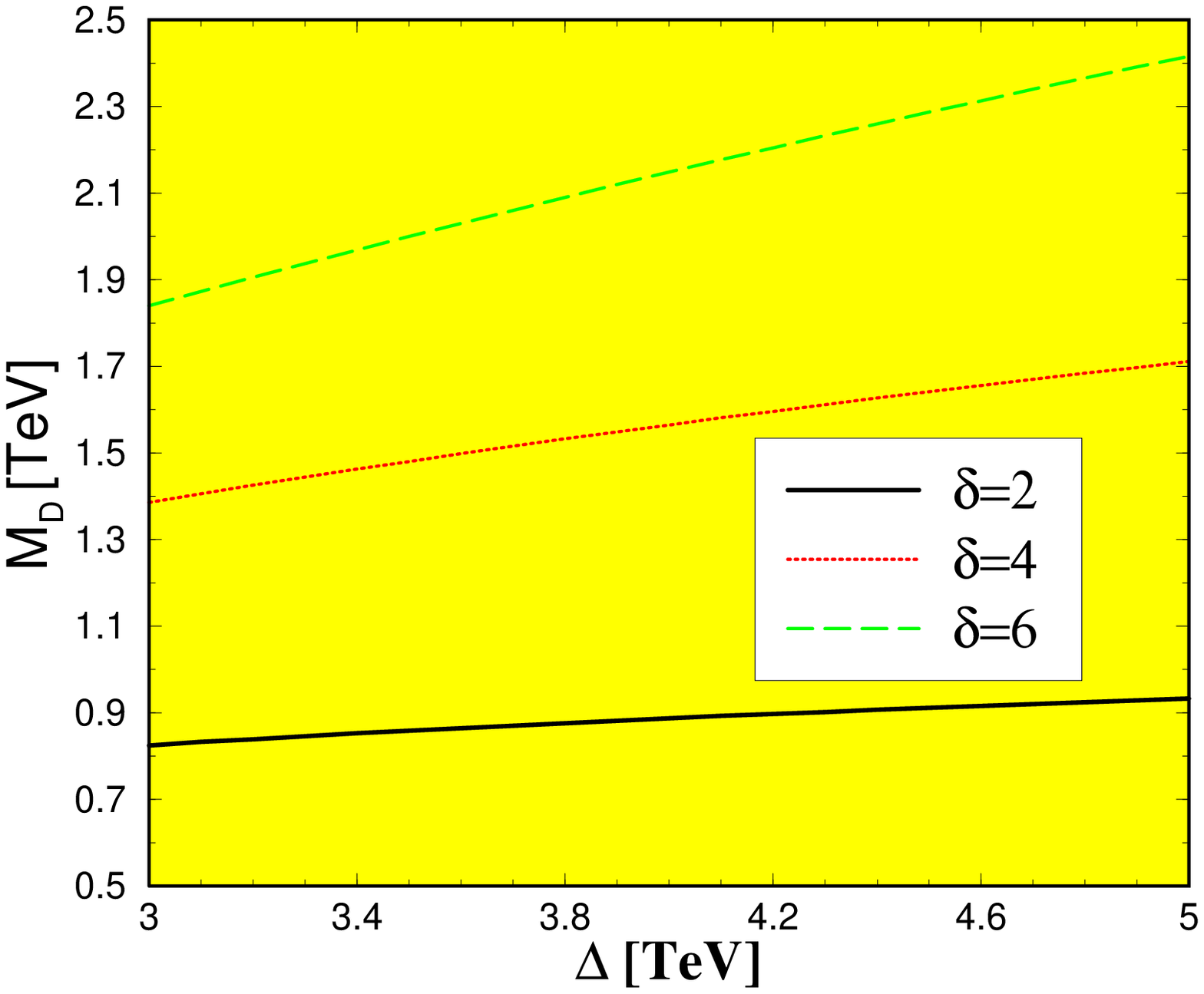,width=6cm,
                    height=6cm}{\it
95 \% C.L. $M_D$ exclusion  bound from the LEPII data with the
process $e^+e^- \rightarrow \gamma\gamma(\gamma)$ with C.M.
energies from 189 to 202 GeV.}{\it 95 \% C.L. exclusion bounds
from the Tevatron data with the processes $p\bar{p} \rightarrow
e^+e^-,\gamma\gamma$ detected by  D0 detector. C.M. energy is
$1.8$ TeV.}

We first consider the LEP-II bound from the $e^+e^- \rightarrow
\gamma\gamma(\gamma)$ \cite{Mele},\cite{Agashe}
 at center of mass
energy ranging from 189 GeV to 202 GeV by DELPHI. The bound was
given as $M_D>713$ GeV with $\lambda=1$ and $M_D>691$ GeV with
$\lambda=-1$ (95\% C.L.)\cite{DELPHI}. As was commented, we  only
consider the case with negative $\lambda$ case. In (Fig.4), the
exclusion bound is presented. For  the case with $\delta=2$, our
new bound is $M_D >810$ GeV and the bound is almost independent of
$\Delta$ scale. Generally the exclusion bound is larger with
larger $\delta$ and increasing with larger $\Delta$.

Now let us consider the Tevatron bound from the diphoton and
dielectron pair production processes\cite{Cheung} obtained by D0.
The center of mass energy is taken about $1.8$ TeV. The lower
limits at 95 \% C.L. on the $M_D$ scale was given between 1.0 and
1.1 TeV with $\lambda=1$ and $-1$ , respectively\cite{D0}. In
(Fig.5), we see the exclusion bound for $M_D$ scale with respect
to the softening scale and number of extra dimensions for the
data. The bound for $\delta=2$ case is similar with LEP-II bound
but in the cases with larger number of extra dimensions the bounds
are much higher. This behavior could be understood by C.M. energy
dependence of $\lambda$ parameter in Eq.(9) and (10).

\section{Summary and conclusion}
We study the implication of the brane fluctuation to the indirect
signals of collider physics. From the exponential softening factor
the brane fluctuation provide very natural regularization scheme
for KK states summation. With this regularization we can determine
$\lambda$ parameter with respect to parameters of $M_D$, $\Delta$
,$\sqrt{s}$ and $\delta$. The experimental exclusion bounds of
existing data and sensitivity bounds for future colliders could be
determined within the parameter space of $(M_D,\Delta,\delta)$. As
a concrete example, we found the exclusion bounds from the LEP-II
diphoton production data with C.M. energy from 189 to 202 GeV and
Tevatron dielectron and diphoton production data with C.M. energy
of 1.8 TeV.

The brane fluctuation provides a new setup for phenomenology of
extra dimension searches. The indirect signals of extra
dimensions, the data is shown to be highly dependent on not only
$M_D$ scale but also $\delta$ and $\Delta$. All the sensitivity
bounds should be determined within the parameter set of $(M_D,
\Delta, \delta)$ if we consider the brane fluctuation.

Finally, let us briefly comment on the brane fluctuation
regularization within the Randall-Sundrum setting \cite{RS}.
Unfortunately, it is not possible to introduce the brane
fluctuation regularization in Randall-Sundrum's case. In the RS
case, our TeV brane carries negative tension to make the warped
geometry and the branes reside on the orbifold fixed points. Thus
we cannot apply any regularization of brane fluctuation. Just
sharp cutoff at reasonable range could provide a finite results of
low energy observables by assuming only KK modes up to the cutoff
scale is responsible for calculation (see e.g.,\cite{Park_RS},
\cite{Agashe_RS}). Further study is needed to formulate natural
regularization formalism with the negative tension brane on the
orbifold fixed point.

\acknowledgments

The work was supported in part by the BK21 program and in part by
the Korea Research Foundation(KRF-2000-D00077).


\begin{thebibliography}{99}
%
\bibitem{ADD}
N.~Arkani-Hamed, S.~Dimopoulos and G.~R.~Dvali, ``The hierarchy
problem and new dimensions at a millimeter,'' Phys.\ Lett.\ B {\bf
429}, 263 (1998) [hep-ph/9803315].
%

\bibitem{Antoniadis:1998ig}
I.~Antoniadis, N.~Arkani-Hamed, S.~Dimopoulos and G.~R.~Dvali,
``New dimensions at a millimeter to a Fermi and superstrings at a
TeV,'' Phys.\ Lett.\ B {\bf 436}, 257 (1998) [hep-ph/9804398].

\bibitem{Han:1998sg}
T.~Han, J.~D.~Lykken and R.~J.~Zhang, ``On Kaluza-Klein states
from large extra dimensions,'' Phys.\ Rev.\ D {\bf 59}, 105006
(1999) [hep-ph/9811350].

\bibitem{Giudice:1998ck}
G.~F.~Giudice, R.~Rattazzi and J.~D.~Wells, ``Quantum gravity and
extra dimensions at high-energy colliders,'' Nucl.\ Phys.\ B {\bf
544}, 3 (1999) [hep-ph/9811291].


%
\bibitem{Sundrum}
R.~Sundrum, ``Effective field theory for a three-brane universe,''
Phys.\ Rev.\ D {\bf 59}, 085009 (1999) [hep-ph/9805471].
%
\bibitem{Strumia}
R.~Contino, L.~Pilo, R.~Rattazzi and A.~Strumia, ``Graviton loops
and brane observables,'' JHEP {\bf 0106}, 005 (2001)
[hep-ph/0103104].
%

\bibitem{Giudice:2003tu}
G.~F.~Giudice and A.~Strumia, ``Constraints on extra-dimensional
theories from virtual-graviton  exchange,'' [hep-ph/0301232].





%
\bibitem{Hewett}
J.~L.~Hewett, ``Indirect collider signals for extra dimensions,''
Phys.\ Rev.\ Lett.\  {\bf 82}, 4765 (1999) [hep-ph/9811356].
%
%
\bibitem{Review}
B.~Olivier, ``Search for Large Extra dimensions at the Tevatron``
[hep-ex/0108015].
%
%
\bibitem{Review2}
M.~Gataullin, ``Searches for Extra Dimensions at LEP``
[hep-ex/0108008].
%
\bibitem{Yan:2001tg}
Q.~S.~Yan and D.~S.~Du, ``Brane fluctuation and the electroweak
chiral Lagrangian,'' Phys.\ Rev.\ D {\bf 65}, 094034 (2002)
[hep-ph/0112236].



\bibitem{Alcaraz:2002iu}
J.~Alcaraz, J.~A.~Cembranos, A.~Dobado and A.~L.~Maroto, ``Limits
on the brane fluctuations mass and on the brane tension scale from
electron positron colliders,'' [hep-ph/0212269].

%
\bibitem{String}
See e.g., E.~Dudas, ``Theory and phenomenology of type I strings
and M-theory,'' Class.\ Quant.\ Grav.\  {\bf 17}, R41 (2000)
[hep-ph/0006190].
%
%
\bibitem{Bando}
M.~Bando, T.~Kugo, T.~Noguchi and K.~Yoshioka, ``Brane fluctuation
and suppression of Kaluza-Klein mode couplings,'' Phys.\ Rev.\
Lett.\  {\bf 83}, 3601 (1999) [hep-ph/9906549].
%
%
\bibitem{Murayama}
H.~Murayama and J.~D.~Wells, ``Graviton emission from a soft
brane,'' [hep-ph/0109004].
%
%

\bibitem{Park}
S.~C.~Park and H.~S.~Song, ``Brane fluctuation and anomalous muon
magnetic moment,'' Phys.\ Lett.\ B {\bf 523}, 161 (2001)
[arXiv:hep-ph/0109121].




%
\bibitem{Mele}
S.~Mele and E.~Sanchez, ``Study of extra space dimensions in
vector boson pair production at LEP,'' Phys.\ Rev.\ D {\bf 61},
117901 (2000) [hep-ph/9909294].
%
%
\bibitem{Agashe}
K.~Agashe and N.~G.~Deshpande, ``Limits on low scale gravity from
e+ e- $\to$ W+ W-, Z Z and gamma gamma,'' Phys.\ Lett.\ B {\bf
456}, 60 (1999) [hep-ph/9902263].
%
%
\bibitem{DELPHI}
P.~Abreu {\it et al.}  [DELPHI Collaboration], ``Determination of
the e+ e- $\to$ gamma gamma (gamma) cross-section at
centre-of-mass energies ranging from 189-GeV to 202-GeV,'' Phys.\
Lett.\ B {\bf 491}, 67 (2000) [hep-ex/0103005].
%
%
\bibitem{Cheung}
K.~Cheung and G.~Landsberg, ``Drell-Yan and diphoton production at
hadron colliders and low scale  gravity model,'' Phys.\ Rev.\ D
{\bf 62}, 076003 (2000) [hep-ph/9909218].
%
%
\bibitem{D0}
B.~Abbott {\it et al.}  [D0 Collaboration], ``Search for large
extra dimensions in dielectron and diphoton  production,'' Phys.\
Rev.\ Lett.\  {\bf 86}, 1156 (2001) [hep-ex/0008065].
%
%
\bibitem{RS}
L.~Randall and R.~Sundrum, ``A large mass hierarchy from a small
extra dimension,'' Phys.\ Rev.\ Lett.\  {\bf 83}, 3370 (1999)
[hep-ph/9905221]
%
%
\bibitem{Park_RS}
S.~C.~Park and H.~S.~Song, ``Muon anomalous magnetic moment and
the stabilized Randall-Sundrum scenario,'' Phys.\ Lett. \ {\bf B}
506, 99 (2001) [hep-ph/0103072]
%
%
\bibitem{Agashe_RS}
K.~Agashe, N.~G.~Deshpande and G.~-H.~Wu, ``Can extra dimensions
accessible to the SM explain the recent measurement of anomalous
magnetic moment of the muon?,'' Phys.\ Lett. \ {\bf B} 511, 85
(2001) [hep-ph/0103235]
%
\end{thebibliography}
\end{document}